%% file: paper.tex
\documentclass[12pt]{article}

\input{newcommands.tex}

\usepackage{psfig}
\usepackage{amssymb}

\begin{document}

\begin{titlepage}
\begin{flushright}
  LU TP 98-15 \\
  hep-ph/9808436 \\
  August 1998
\end{flushright}
\vspace{25mm}
\begin{center}
  \Large
  {\bf The Feynman--Wilson gas and the Lund model} \\
  \normalsize
  \vspace{12mm}
  B.~Andersson, G.~Gustafson, M.~Ringn\'{e}r and
  P.J.~Sutton\footnote{bo@thep.lu.se, gosta@thep.lu.se,
	markus@thep.lu.se, peter@thep.lu.se}
  \vspace{3ex} \\
  Department of Theoretical Physics, Lund University, \\
  S\"olvegatan 14A, S-223 62 Lund, Sweden \\
\end{center}
\vspace{25mm}

\noindent We derive a partition function for the Lund fragmentation
model and compare it with that of a classical gas.  For a fixed
rapidity ``volume'' this partition function corresponds to a
multiplicity distribution which is very close to a binomial
distribution. We compare our results with the multiplicity
distributions obtained from the JETSET Monte Carlo for several
scenarios. Firstly, for the fragmentation vertices of the Lund
string. Secondly, for the final state particles both with and without
decays.

\vspace{3cm}

\end{titlepage}

\section{Introduction} \vspace{-2ex}
\input{intro.tex}

\section{The Feynman--Wilson gas}
\input{fwg.tex}

\section{The Lund model and the Feynman--Wilson gas}
\subsection{The Lund model}
\input{lund.tex}

\subsection{The connection between the Lund model and the FWG}
\input{lund_fwg.tex}

\section{The vertex distributions}
\input{vert.tex}

\section{The particle distributions}
\input{part.tex}

\section {Conclusions}
\input{conclusions.tex}

\vspace{1cm} {\noindent \large \bf Acknowledgment}\vspace{0.5cm} \\
This work was supported in part by the EU Fourth Framework Programme
`Training and Mobility of Researchers', Network `Quantum Chromodynamics
and the Deep Structure of Elementary Particles', contract
FMRX-CT98-0194 (DG 12 - MIHT).

\appendix
\section{The binomial and negative binomial distributions}
\input{binomial.tex}

\section{The binomial approximation of the partition function}
\input{lund_binomial.tex}

\input{references.tex}
\end{document}

%% file: newcommands.tex
\newcommand{\bibl}[5]
	{#1, {\it #2} {{\bf #3},} #5 (#4)}
\newcommand{\cebe}{\begin{center}}
\newcommand{\ceen}{\end{center}}
\newcommand{\debe}{\begin{description} \vspace{-2ex}}
\newcommand{\deen}{\end{description}}
\newcommand{\eabe}{\begin{eqnarray}}
\newcommand{\eaen}{\end{eqnarray}}

\newcommand{\eqbe}{\begin{equation}}
\newcommand{\eqen}{\end{equation}}

\newcommand{\itbe}{\begin{itemize}}
\newcommand{\iten}{\end{itemize}}

\newcommand{\tabe}{\begin{tabbing}}
\newcommand{\taen}{\end{tabbing}}

%% file: intro.tex
The cross-sections of QCD multiparticle production processes at high
energies have many similarites with the multiparticle distributions of
a classical gas, an analogy which was first noted by Feynman and
Wilson~\cite{r:Wilson}. This gas is essentially one dimensional in
rapidity space. In this paper we use the gas analogy to derive a
partition function for the Lund string fragmentation
model~\cite{r:lund}. We perform a virial expansion to the second order
in the density of particles. Our partition function then yields an
equation of state for a Van der Waal's gas. Furthermore, it reduces to
that of an ideal gas when the produced particles are massless.

The partition function of the gas is related in a simple way to the
multiplicity distribution of its constituent particles. This provides
us with a method of investigating the partition function.  We show
that for a fixed rapidity ``volume'' our partition function
corresponds to a multiplicity distribution which is very similar to a
binomial distribution.

For large rapidity intervals the major fluctuations in multiplicity
stem from gluon radiation. We will, however, neglect gluon
emission. In this paper we are only interested in comparing the Lund
fragmentation model with the properties of a classical gas.

We analyse the multiplicity distributions obtained from the JETSET
Monte Carlo~\cite{r:jetset} for several scenarios. Firstly, we
investigate the string break-up vertices, then the primary particles
and finally we include decays. We find that all cases are remarkably
well described by distributions from the binomial family. In the
derivation of our partition function we assume that the particles are
ordered in rapidity. Since this is true for the vertices, we expect
the distributions of vertices to be the optimal case. Indeed, these
distributions are well described by our partition function.

The transition from vertices to particles introduces some smearing
in rapidity. This results in a wider multiplicity distribution, where
the width is sensitive to the transverse mass of the produced
particles. We obtain an ordinary binomial for the primary
particles. However, the strong smearing from decays ensures that, for
the final state particles, this distribution becomes a negative
binomial distribution.

We shall begin with a short presentation of the basic ideas of the
Feynman--Wilson gas~(FWG). This is followed by an introduction to the Lund
model and its relationship to the FWG. We next turn to the
multiplicity distributions for the vertices and lastly how they
are modified for the final state particles.

%% file: fwg.tex
\label{s:fwg} 
The original discussion of the FWG can be found in
\cite{r:Wilson}. Here we summarize the main features of the model.
We consider a multiparticle production process where the two primary
particles have four momenta $p_1$ and $p_2$ and large invariant
$s=(p_1+p_2)^2$. The $n$ secondary particles have four momenta
$k_1, k_2, \dots, k_n$, and each is on the mass shell.  In the FWG
model the three remaining degrees of freedom in each $k_i$ correspond
to the ``spatial'' co-ordinates of a gas particle via
\begin{eqnarray}
\tilde{x} &=& k_x \nonumber \\
\tilde{y} &=& k_y \nonumber \\
\tilde{z} &=& \ln[(k_z+k_0)/m_\perp]
\label{eqnb}
\end{eqnarray}
where the transverse mass is defined by
\begin{equation}
m_\perp = \sqrt{m^2+k_x^2+k_y^2} \; \; .
\end{equation}
Note that in this picture $\tilde{z}$ corresponds to
the rapidity of the relevant particle. 
We will assume here that each produced particle is of the same
type (each has the same mass)
but the extension to different species is straightforward.

We can write the total cross section for the production process
using these spatial variables. We first note that
the invariant phase space $d^3 k /k_0$ becomes
$d^3 \tilde{r}$. The energy momentum conserving delta functions
are first written in terms of $p = p_1+p_2-k_1- \dots -k_n$.
\begin{equation}
\delta (p_0) \delta^3 (p) = 2\delta (p_+) \delta (p_-) \delta^2 (p_\perp)
\end{equation}
with $p_\pm=p_0 \pm p_z$. This can be expressed in terms of
$\tilde{r}$ variables using the relationship $k_0 \pm k_z = m_\perp
e^{\pm \tilde{z}}$.

The delta functions have the effect of introducing a fixed volume for
the gas.  The transverse momenta are limited and constrain the gas to
a narrow tube of radius $\sim 300$ MeV. We shall instead focus on
the $\tilde{z}$ co-ordinate.  We first introduce $W_+$ and $W_-$ via
\begin{equation}
W_\pm \equiv (p_1+p_2)_\pm 
\end{equation}
so that we can write
\begin{equation}
\delta (p_\pm)=\delta(W_\pm-\sum m_{\perp i} \exp (\pm \tilde{z}_i))\;.
\end{equation}
In the following we use the Lorentz frame where $W_\pm=\sqrt{s}$.  The
two delta distributions contain the requirement that the the ``gas
volume'' should be of the order of $\ln s$. To see this we may
integrate out the rapidities of the first and the last particles to
obtain
\begin{eqnarray} 
& & d\tilde{z}_1d\tilde{z}_n \,\delta(\cdots) \delta (\cdots) \simeq 1/s  \nonumber  \\ 
& & \tilde{z}_1 \simeq - \tilde{z}_n \; \simeq \; \ln(\sqrt{s}) \;.
\end{eqnarray} 
We may in this approximation choose a
number $s_0$ in such a way that  
\begin{equation} 
\Delta \tilde{z} \equiv \tilde{z}_1 - \tilde{z}_n = \ln(s/s_0)
\end{equation}  
and assume that all the particles are kept inside this rapidity
``volume''. If the spatial co-ordinates of the primary particles are
$R_1$ and $R_2$ respectively then the cross section can be written as
\begin{equation}
\sigma_T(R_1,R_2) = \sum_{n=2}^{\infty}  \left[\left(\prod_{i}^{n} 
\int d^3 \tilde{r}_i\right) \; 2 \delta (p_+) \delta(p_-) \delta^2(p_\perp) \;
\sigma_n(\tilde{r}_1, \dots , \tilde{r}_n,R_1,R_2) \right ] \;.
\end{equation}
For fixed $R_1, R_2$ then $\sigma_T$ corresponds, in the FWG analogy,
to the partition function of the gas and the functions $\sigma_n$ are
the $n$ particle distribution functions for the gas.  Our aim is to
connect these ideas to particle production within QCD as represented
by the Lund model.

%% file: lund.tex
\label{s:lund}
In this section we briefly review some features of the Lund model
fragmentation scheme. We will mostly be concerned with the simple
situation when the colour force field from an original
quark-antiquark pair (produced by $\mbox{e}^+\mbox{e}^-$~annihilation, 
for example) decays into a set of final state hadrons.

In the Lund model, the colour force field is approximated by a massless
relativistic string with a quark ($\mbox{q}$) and an antiquark 
($\overline{\mbox{q}}$) at
the endpoints. The gluons are treated as internal excitations on the
string field. This means that there is a constant force field, $\kappa
\simeq 1~\mbox{GeV}/\mbox{fm}$, corresponding to a linearly rising
potential, spanned between the original pair. After being produced the
$\mbox{q}$ and the $\overline{\mbox{q}}$ are moving apart and the
energy in the field can be used to produce new
$\mbox{q}\overline{\mbox{q}}$-pairs. When a new pair is created the
string is split into two pieces.

\begin{figure}[t]
  \hbox{\vbox{ \begin{center}
    \mbox{\psfig{figure=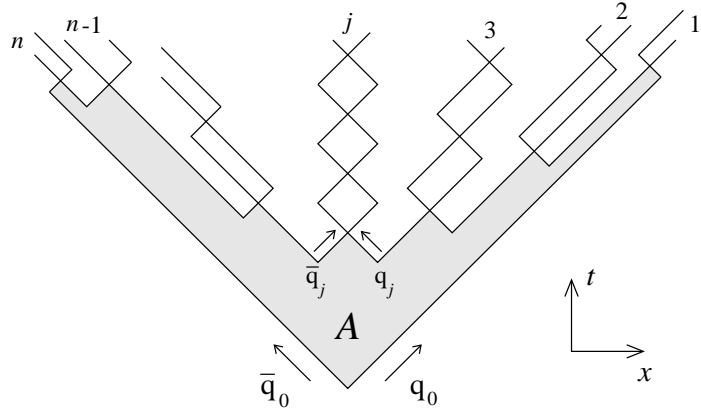,width=9.2cm}} 
	\end{center} }}
\caption{\small The break-up in space--time of a Lund string into $n$ hadrons.
The fragmentation area is denoted by $A$.}
\label{f:breakup}
\end{figure}

The production rate of a pair with combined
internal quantum numbers corresponding to the vacuum is, from quantum
mechanical tunneling in a constant force field, given by
\begin{equation}
\label{e:tunneling}
P(\mu_\perp) = \exp\left(-\frac{\pi \mu_{\perp}^2}{\kappa}\right).
\end{equation}
Here the quarks in the pair have transverse mass
$\mu_{\perp}=\sqrt{\mu^2+\vec{k}_{\perp}^2}$, mass $\mu$ and
transverse momentum $\pm\vec{k}_{\perp}$.  The final state mesons in
the Lund model correspond to isolated string pieces containing a
$\mbox{q}$ from one breakup vertex and a $\overline{\mbox{q}}$ from
the adjacent vertex together with the produced transverse momentum and
the field energy in between. The break-up of the string is illustrated
in Fig.(\ref{f:breakup}).

One necessary requirement is that to obtain real positive (transverse)
masses all the vertices must have spacelike difference
vectors. Together with Lorentz invariance this means that all the
vertices in the production process must be treated in the same way
\cite{r:bs83}. Another consequence is that it is always the slowest
mesons that are produced first in any Lorentz frame (corresponding to
the fact that time-ordering is frame dependent). Furthermore each
vertex has the property that it will divide the event into two
causally disconnected jets, the mesons produced along the string field
to the right and those produced to the left of the vertex.  This can
be seen in Fig.(\ref{f:breakup}).

A convenient ordering along the force field of the produced particles
is rank ordering. Two particles have adjacent rank if they share a
$\mbox{q}\overline{\mbox{q}}$ pair created at a vertex . The first
rank meson contains the internal quantum numbers of the original
$\mbox{q}$ together with those of the $\overline{\mbox{q}}$ produced
at the vertex closest to the endpoint $\mbox{q}$. Similarly
the second rank meson contains the internal quantum numbers of the
$\mbox{q}$ from this ``first'' vertex and the $\overline{\mbox{q}}$ of
the ``second'' etc. In this way rank ordering corresponds to an
ordering along a light-cone. Alternatively it is also possible to rank order
in the direction from the original $\overline{\mbox{q}}$.

The basic Lund model fragmentation process then stems from the
following two assumptions

\begin{enumerate}

\item In the centre of phase space (i.e. far from the endpoints) the string
decay process will reach a steady state. The probability to find a
vertex is, after many production steps along the light-cone, a finite
distribution in the proper time of the vertex. This is also the case
when the total string field energy becomes very large.

\item The decay process is the same whether it is 
ordered along the positive or along the negative light-cone. 

\end{enumerate}

\begin{figure}[t]
  \hbox{\vbox{ \begin{center}
    \mbox{\psfig{figure=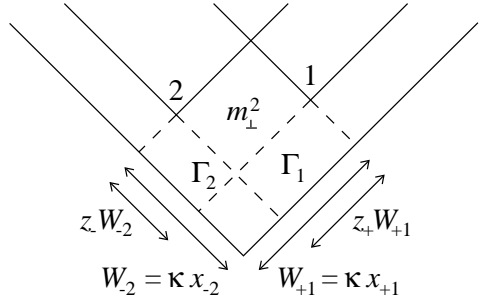,width=6.3cm}} 
	\end{center} }}
\caption{\small The production, in energy--momentum space, of 
a particle with transverse mass $m_\perp$. The particle is produced
between the vertices 1, with the squared proper-time
$\tau_1^2=\Gamma_1/\kappa^2$, and 2, with
$\tau_2^2=\Gamma_2/\kappa^2$.  The particle has fractional light-cone
components $z_+$ and $z_-$. }
\label{f:particleprod}
\end{figure}

If we consider Fig.(\ref{f:particleprod}), this means that we assume
that the probability to reach the space-time point 1 at
$(x_{+1},x_{-1})$, after many steps along the positive light-cone, and
to produce a meson with transverse mass $m_{\perp}$ by one further
step to the vertex 2 at $(x_{+2},x_{-2})$, is equal to the probability
to reach the point 2, after many steps along the negative light-cone,
and by one further step to 1 produce the meson with $m_{\perp}$.
   
Changing variables to the squared Lorenz invariant proper time
$\tau^2=x_{+}x_{-}$ and the rapidity $y=1/2\ln{(x_{+}/x_{-})}$ the
probability to reach the point 1 is $H(\tau_1^2) d\tau_1^2
dy_1$. The probability to produce one further particle with mass
$m_{\perp}$ and fractional light-cone component $z_+$ is
$f(z_+,m_{\perp})dz_+$.  A particle with fractional light-cone
component $z_+$ has the positive light-cone energy-momentum component
$p_+= z_+ \kappa x_{+1}$ and has, in order to stay on the mass-shell,
the negative component $p_- = m_{\perp}^2/p_+ = z_- \kappa
x_{-2}$. This means that we obtain the equation:
\begin{equation}
\label{e:Lundass}
H(\tau_1^2)d\tau_1^2 f(z_+,m_{\perp})dz_+ = H(\tau_2^2)d\tau_2^2
f(z_-,m_{\perp}) dz_- \;.
\end{equation}
It is a nice and surprising feature of the assumptions above that
there is a unique process that fulfills
Eq.(\ref{e:Lundass})~\cite{r:bs83},
\begin{eqnarray}
\label{e:Hfres}
H_j &=& C_j\Gamma^{a_j} \exp(-b\Gamma)~~~~\mbox{with}~~ \Gamma= \kappa^2 \tau^2 \;, \nonumber \\
f_{jk}&=& \hat{N}_{jk}z^{a_j-1}\left(\frac{1-z}{z}\right)^{a_k} \exp(-bm_{\perp}^2/z) \;. 
\end{eqnarray}
The numbers $C_j$ and $\hat{N}_{jk}$ are normalisation constants and
the particle is assumed to be produced in a step from a vertex with
flavour $j$ to a vertex with flavour $k$. If $n_f$ denotes the number
of $\mbox{q}\overline{\mbox{q}}$-flavours, the process has $n_f+1$
parameters.  Although the parameter $a$ is, in principle, flavour
dependent, there has been no need to utilize this in the Lund model as
implemented in the JETSET Monte Carlo; except for the first rank
particle in a heavy quark jet~\cite{r:Bowler}. The
parameter $b$ must be flavour independent.

\begin{figure}[t]
  \hbox{\vbox{ \begin{center}
    \mbox{\psfig{figure=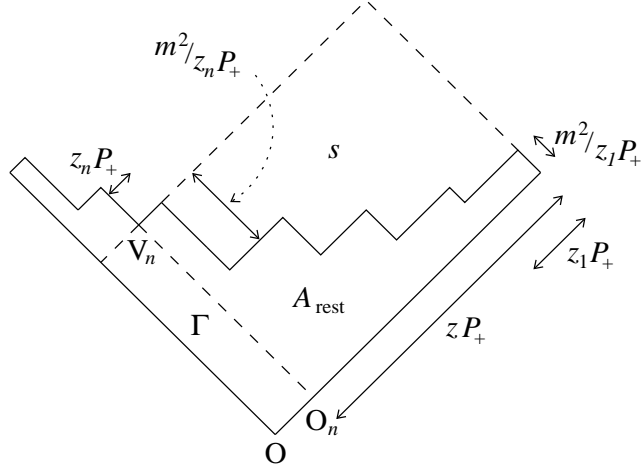,width=8.5cm}} 
	\end{center} }}
\caption{\small The decay, in energy-momentum space, of an $n$-particle cluster with invariant squared mass $s$.
The fragmentation area of the cluster is $A_{\mbox{\scriptsize
rest}}$. $\Gamma=\kappa^2\tau^2$ is with respect to the proper-time
$\tau$ of the last vertex.}
\label{f:cluster}
\end{figure}

It is possible to construct the probability to
produce a finite energy cluster of rank-connected
particles~\cite{r:bs83} from Eq.(\ref{e:Hfres}). Such a cluster 
is shown in Fig.(\ref{f:cluster}). 
This probability distribution is in a natural way subdivided into two
parts, the probability to obtain the cluster and the probability that
the cluster decays in a particular way. In the following we order the
particles along the positive light-cone. If the cluster has a total
light-cone fraction $z$ and a fixed total squared cms energy~$s$ then the
(non-normalised) probability to obtain such a cluster is
\begin{equation}
\label{e:external}
dP_{\mbox{\scriptsize ext}} =
\frac{dz}{z} z^{a_0} \left(\frac{1-z}{z}\right)^{a_n}\exp(-b\Gamma(s,z)) 
~~~~\mbox{with}~~  \Gamma(s,z) = s\frac{1-z}{z} \;.
\end{equation}
The cluster then starts at a vertex with flavour $f_0$ and ends with
flavour $f_n$. The $\Gamma$ value is that of the last vertex,
as shown in Fig.(\ref{f:cluster}). 
Thus a cluster is produced in the same way as a single particle 
between the vertices with $a_0$ and $a_n$.
Similarly we find that the (non-normalised)
probability for the cluster to decay into the particular channel with
the particles $\{p\}_j$ is
\begin{equation}
\label{e:internal}
dP_{\mbox{\scriptsize int}}= \left[\prod \hat{N}_j dp_j 
\delta(p_j^2 - m_j^2)\right] \delta (\sum p _j -P_{\mbox{\scriptsize tot}}) 
\exp(-bA_{\mbox{\scriptsize rest}}) 
\end{equation}
where $A_{\mbox{\scriptsize rest}}$ is the decay area of the cluster,
as shown in Fig.(\ref{f:cluster}). Equation~(\ref{e:internal}) is for
simplicity written in the ordinary Lund model fashion with a single
$a$-parameter (this parameter is not explicit in the
formula) and we note the appearance of the phase space for the final
state particles multiplied by the exponential area decay law. The
quantity $P_{\mbox{\scriptsize tot}}$ is the total energy momentum of
the cluster so that $P_{\mbox{\scriptsize tot}}^2=s$. We may 
determine the finite energy version of the vertex distribution, $H(\Gamma)$,
from Eq.(\ref{e:external}) by exchanging $z$ for $\Gamma$. This yields
\begin{eqnarray}
\label{e:finiteH}
H_s \propto \frac{\Gamma^{a_n}s^{a_0-a_n}}{(\Gamma+s)^{a_0+1}}
\exp(-b\Gamma) \; .
\end{eqnarray}
The function $H_s$ in Eq.(\ref{e:finiteH}) is exponentially decreasing
in $\Gamma$ so that the power dependence in the denominator only plays
a role for small values of $\Gamma$ and then it is hardly noticable
for large values of $s$.  In this way the assumption $1.$ above is
fulfilled. That is to say when $s$ becomes very large there is (after
normalisation) a finite distribution in the proper-time size of the
decay vertices.

%% file: lund_fwg.tex
\begin{figure}[t]
  \hbox{\vbox{ \begin{center}
    \mbox{\psfig{figure=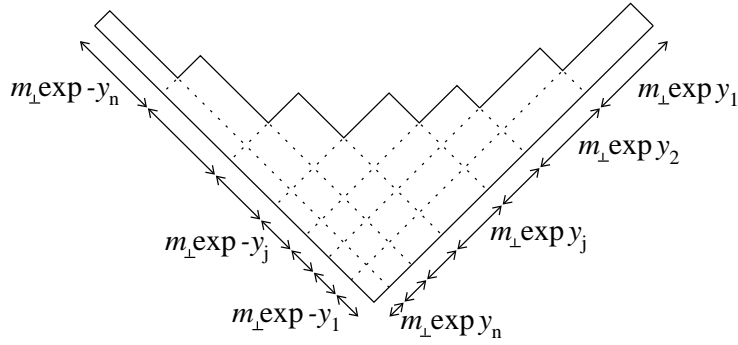,width=9.7cm}} 
	\end{center} }}
\caption{\small The fragmentation area partitioned into 
two-particle regions reveals how the
area can be expressed as in Eq.(\ref{e:area1}).}
\label{f:area_partition}
\end{figure}

\label{s:lund_fwg}
We will now exhibit the decay distribution of a cluster, as given by
Eq.(\ref{e:internal}), in terms of the partition function which is
studied in statistical physics. For simplicity we write the formulas
for a single particle transverse mass $m_{\perp}$ and a single flavour and we
let $j$ denote the rank of a particle. The phase space factor can in
analogy with the result in section~\ref{s:fwg} be written with the
particle energy momentum vectors $p_j \equiv m_{\perp} (\exp(y_j),
\exp(-y_j))$ as
\begin{eqnarray}
\label{e:phasespace} 
d\Psi &\equiv& \left[\prod \hat{N} dp_j \delta(p_j^2 - m_{j}^2) \right] 
\delta (\sum p_j -P_{tot}) \nonumber \\
&=& 
\left[\prod_1^n \hat{N} dy_j \right]
\delta (\sum m_{\perp}\exp(y_j) -P_{+})
\delta (\sum m_{\perp}\exp(-y_j)-P_{-}) \nonumber \\
&\simeq& 
\frac{\hat{N}^2}{s}\prod_2^{n-1} \hat{N} dy_j \;. 
\end{eqnarray}
We have in the last line integrated out the first and the last
rapidities in the delta function and from now on we assume that the
remaining particles are placed in rapidities between $\Delta y/2$ and 
$-\Delta y/2$ with $\Delta y=\ln(s/s_0)$ and $s_0$ is some suitable
scale. If all the particles are ordered in rapidity we may integrate
out the phase space factor and obtain
\begin{equation}
\label{e:phase2}
\int d\Psi = \frac{\hat{N}^2(\hat{N} \Delta y)^{n-2}}{s(n-2)!} \; .
\end{equation}
We next consider the decay area of the cluster. 
Figure~(\ref{f:area_partition}) shows
that it can be written
in terms of the rapidities of the particles
\begin{equation}
\label{e:area1}
A=m_{\perp}^2 \sum_{j=1}^n\sum_{k=j}^n \exp(y_k-y_j)\; .
\end{equation}
From this equation and Eq.(\ref{e:phasespace})
we note that the decay distribution in Eq.(\ref{e:internal}) has
similarities with a partition function, $Z_n$, and we therefore define
a grand partition function $Z$ as
\begin{eqnarray}
Z = \sum_n Z_n & = & s\;\sum_n 
\left[ \left( \prod_{j=1}^n \hat{N} dy_j \right)\delta(\ldots)\delta(\ldots) 
\exp\left(-bm_{\perp}^2 \sum_{j=1}^n \sum_{k=j}^n 
\exp(y_k-y_j)\right) \right] \nonumber \\
& \equiv  & s\; \sum_n \left[ \left( \prod_{j=1}^n \hat{N} dy_j \right) 
\delta(\ldots)\delta(\ldots)\exp\left(-\frac{1}{kT}
\sum_{j=1}^n \sum_{k=j}^n V(y_j-y_k)\right) \right] \;.
\label{e:Z}
\end{eqnarray}
(The factor of $s$ is required in order to have a dimensionless partition
function.) In this way we see that the decay distribution in
Eq.(\ref{e:internal}) may be interpreted as the partition function for
a system of $n$ particles with co-ordinates $y_j$ interacting with
exponential two-body potentials in a one-dimensional volume equal to
$\Delta y$.  We note that whilst all the particles interact in this
way (``long-range interactions'') the exponential decrease of the
potentials ensures that the effective interaction is rather short
ranged.

If the particles are imagined as making up a gas
in rapidity space and are interacting via two-body potentials, then
the Hamiltonian is
\begin{equation}
H=\sum_j T(\pi_j)+\sum_{j,k} V(y_j-y_k) \; .
\end{equation}
The phase space volume element is $\prod (dy_jd\pi_j)$, with
$\pi_j$ denoting the quantities canonically conjugate to the
co-ordinates $y_j$. The kinetic energy factors $T$ are integrated
out in Eq.(\ref{e:Z}) and incorporated into the constants $\hat N$. These
constants then play the role of fugacities.  

We shall now attempt to obtain a simplifed expression for
the partition function, $Z_n$. 
We have already seen that in the strong ordering
limit the phase space may be easily integrated according to
Eq.(\ref{e:phase2}). Using this limit may seem drastic since two
neighbours in rank may well have a different rapidity order. However,
if many pairs are not well ordered then many of the exponential
potentials in the in the partition function will be strongly
increasing, i.e. the area suppression in the Lund model will make
these contributions small.

We can easily find an expression for the exponential in our partition
function if we approximate the fragmentation area.  Assuming that each
particle lies along the hyperbola with $<\sqrt{\Gamma}>=\gamma$ then
each takes up a rapidity length $\delta =
m_\perp/\gamma$. Consequently all the $n$ particles take up the
rapidity length $\Delta y=n\delta = n m_\perp/\gamma$ and the total
area is $\gamma^2 \Delta y = n^2 m_\perp^2/\Delta y$. Since the
particles are produced around the average hyperbola, we expect that
this result may be modified by a constant, $c_2$, of order unity
giving
\begin{eqnarray}
\label{e:negaarea}
bA\simeq \frac{bc_2  m_\perp^2 n^2}{\Delta y} = \frac{c_3n^2}{\Delta y}\;,  
\end{eqnarray}
where we have introduced $c_3=bm_\perp^2c_2$.
In this way we obtain from Eq.(\ref{e:phase2}) and
Eq.(\ref{e:negaarea}) a description of the grand partition function in
terms of the multiplicity $n$. In the approximation that $n$ is large, i.e.
for large rapidity intervals $\Delta y$, we can write the partition function 
in terms of two parameters $c_1$ and $c_3$ as
\begin{equation}
Z_n \simeq \frac{(c_1 \Delta y)^n}{n!}\exp\left(\frac{- c_3 n^2}{\Delta y} \right)\;.
\label{e:Z_n}
\end{equation} 
We will comment further on the parameters $c_1$ and $c_3$ when we
investigate to what extent the partition function in Eq.(\ref{e:Z_n})
describes the particle production in the Lund model.

\subsection{The partition function in the Gaussian approximation}
\label{s:lund_fwg_bo}

We now investigate the grand partition function in the limit where the
number of particles is large, but the density is low (as in an
ordinary gas).  In this case we expect that the grand partition
function can be approximated by the maximal term in the sum.  To find
the multiplicity for which the partition function is maximal we first
define $\Phi_n$ by writing Eq.(\ref{e:Z_n}) as
\begin{equation}
Z_n = \exp \Phi_n \; .
\end{equation}
If we treat $n$ as a continuous variable we
can expand $\Phi_n$ in a Taylor series as
\begin{equation}
\Phi(n) \simeq \Phi({\overline n}) + (n-\overline n) \;
\Phi^{\prime}({\overline n})
+ \frac{(n-\overline n)^2}{2} \;
\Phi^{\prime \prime}({\overline n})
\; .
\end{equation}
Choosing ${\overline n}$ such that $\Phi^{\prime}({\overline n})=0$,
we evidently have a Gaussian approximation for $Z_n$
\begin{equation}
Z_n \simeq \exp \Phi({\overline n}) \; 
\exp \left ( - \frac{(n-{\overline n})^2}{2 V} \right ) 
\end{equation}
where the variance, $V$, is given 
by $V=1/\Phi^{\prime \prime}({\overline n})$.
It is straightforward to obtain expressions for both 
${\overline n}$ and $V$ if we use 
Stirlings approximation for the factorial
in $\Phi(n)$. We find 
\begin{eqnarray}
{\overline n} 
&=& \frac{\Delta y}{2 c_3} \; \ln(\frac{c_1 \Delta y}{\overline n}) \nonumber \\
V&=& {\overline n} \left ( 1+ 
\frac{2 c_3 {\overline n}}{\Delta y} \right )^{-1} \; .
\label{e:mult_variance}
\end{eqnarray}
Notice that, since $c_3$ is positive, this implies that the
variance of the distribution is less than the mean and
the distribution is therefore narrower than a Poissonian.
If we now introduce the density of particles in the rapidity volume,
$R=\overline{n}/\Delta y$, then
\begin{equation}
\Phi({\overline n}) = (R+c_3R^2)\Delta y \;.
\end{equation}
For large ${\overline n}$ we can approximate the grand partition
function as $Z \sim Z_{\overline n}$ and so
\begin{equation}
Z \sim \left(\frac{s}{s_0}\right)^{a_R}
\label{e:grand_Z}
\end{equation}
with 
\begin{equation}
a_R={R+c_3R^2} \; .
\label{e:a_eff}
\end{equation}

The grand canonical partition function,
for a gas is related to 
the pressure, $P$, temperature, $T$ and volume, $\ln(s/s_0)$,
of the gas via
\begin{eqnarray}
\Omega &\equiv&-kT\ln Z \nonumber \\
P&=&-\frac{\partial \Omega }{\partial \ln(s/s_0)} 
\end{eqnarray}
where $k$ is Boltzmann's constant. 
For the partition function in Eq.(\ref{e:grand_Z}) 
we obtain the following equation of state for
the gas
\begin{equation}
P=kT(R+c_3R^2) \;.
\label{e:eqn_state}
\end{equation}
Our expansion thus corresponds to the first two terms in the virial
expansion in the particle density of the gas. We note that the
equation of state in Eq.(\ref{e:eqn_state}) is similar to that of a
Van der Waal's gas. For particles with zero (transverse) mass we have
$c_3=0$. In this case a particle does not take up any volume in
rapidity and Eq.(\ref{e:eqn_state}) reduces to the equation of state
for an ideal gas.

%% file: vert.tex
\label{s:vertex}
 
The partition function is related to the multiplicity
distribution, $P_n$, since
\begin{equation}
P_n \; = \frac{Z_n}{Z} \; \; .
\end{equation}
In the remaining sections we shall use this relationship to further study our
partition function. We begin here with a study of the
vertices produced in the string fragmentation. These vertices
are strongly ordered in rapidity and thus satisfy one of the
assumptions used to derive our partition function. This is
only an approximation in the case of the particles. Of course,
the number of vertices corresponds directly to the number of primary particles.

In what follows we outline a simple model in which all particles have
the same mass ($m=0.8$ GeV) and there is no transverse momenta. The
effects of relaxing those constraints will be considered in the next
section where we return to the particles.

\subsection{The distribution in rapidity}

\begin{figure}[t]
  \hbox{\vbox{ \begin{center}
    \mbox{\psfig{figure=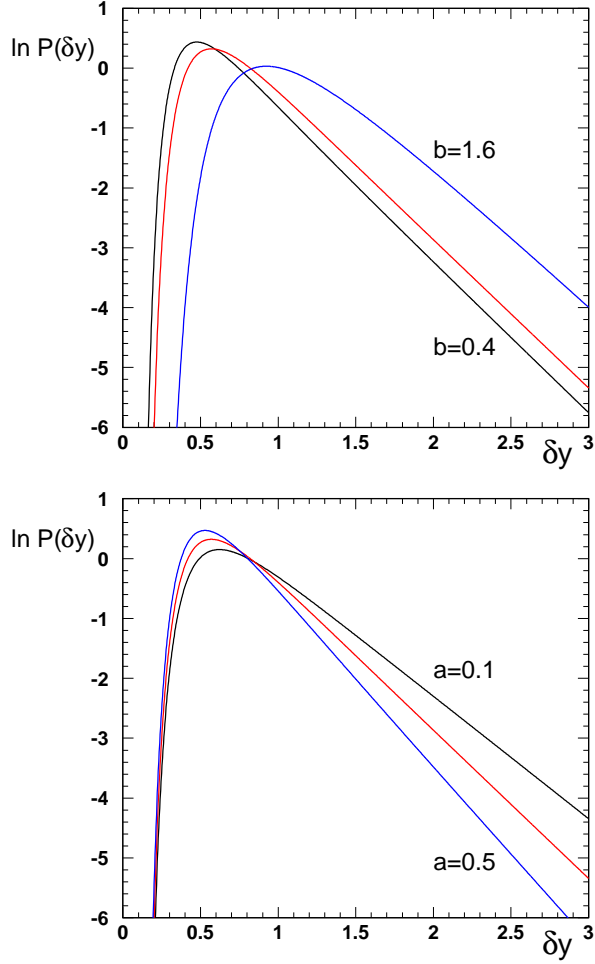,width=10cm}} 
\end{center} }}
\caption{\small 
The logarithm of the distribution of the rapidity distance $P(\delta y)$ between
adjacent vertices as predicted by the Lund model
(Eq.(\ref{e:step_lund})) for a fixed mass, $m=0.8$ GeV, but different
values of the Lund model parameters $a$ and $b$. The upper plot shows
fixed $a=0.3$ and $b=0.4$, 0.58, 1.6.  The lower plot shows fixed
$b=0.58$ and $a=0.1$, 0.3, 0.5. }
\label{f:lund_vert}
\end{figure}

We begin by studying the separation between neighbouring vertices. In
the Lund model the distribution of such separations for a fixed mass,
$m$, is given by
\begin{equation}
P(\delta y) = N \int d \Gamma \; \Gamma^a e^{-b \Gamma} \; \int_0^1 dz \;
\; \frac{(1-z)^a}{z} \; e^{-bm^2/z}  \; \delta
\left ( 
\delta y - \frac{1}{2} \ln \left ( \frac{\Gamma+m^2/z}{\Gamma (1-z)} \right )
\right ) \; .
\label{e:step_lund}
\end{equation}
The logarithm of this distribution is plotted in Fig.(\ref{f:lund_vert}) 
for various values of the Lund model parameters $a$ and $b$.
We see from this figure that there are two main characteristics of the
distribution. The first is an effective minimum separation between
vertices which increases with $b m^2$, but is independent of $a$. 
Physically this separation arises because two
vertices cannot be very close together in rapidity if
they must produce a massive particle.
The second characteristic is an
exponential fall off for large separations, $\delta y$, which 
depends only on the parameter $a$.

We can consider a simple model which reproduces the above features
very well. In this model the rapidity region is divided up into a
series of $N$ equal bins of size $\delta y_{\mbox{\scriptsize
bin}}$.  The effective minimum separation between vertices can now 
be taken into account by demanding that no bin may contain more than a
single vertex. Each bin is assigned a
probability $p$ to contain a vertex and a probability $1-p$ to be
empty. This allows us to compute the probability of
a separation, $\delta y$, between two vertices.
If $\delta y$ is discretised as $\delta y = n \delta
y_{\mbox{\scriptsize bin}}$ with $n$ an integer then the probability of
such a separation is given by a geometric series
\begin{eqnarray}
P(\delta y )&=&p(1-p)^{n - 1} ~~~~~~~~~~~~~~~ (n = 1, 2, \dots) \nonumber \\
&=&\frac{p}{(1-p)} \exp (-\beta \delta y)
\end{eqnarray}
with $\beta= -\ln(1-p)/\delta y_{\mbox{\scriptsize bin}}$.
We see that large $\delta y$ separations are exponentially
suppressed. 
The two main features of Fig.(\ref{f:lund_vert}) 
are thus very well reproduced by this simple model, which
corresponds to distributing the vertices
according to a binomial distribution (appendix \ref{s:binomial}). 

\begin{figure}[t]
  \hbox{\vbox{ \begin{center}
    \mbox{\psfig{figure=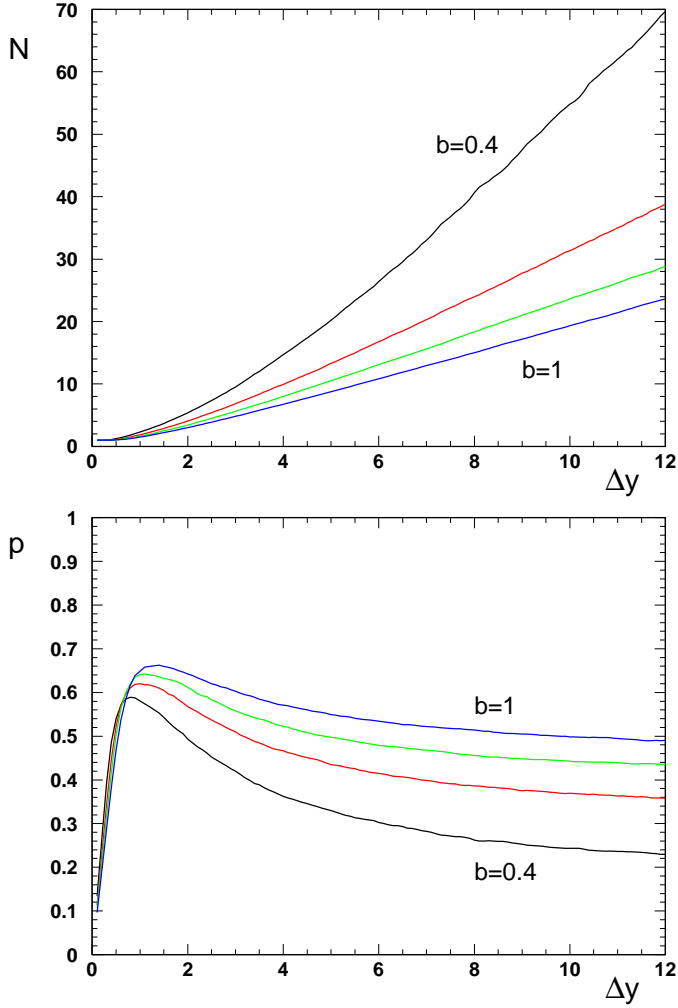,width=10cm}} 
\end{center} }}
\caption{\small The values of $N$ (upper plot) 
and $p$ (lower plot) for the vertex distributions 
produced by JETSET (for a fixed mass and 
no transverse momentum generation) 
as a function of the rapidity volume, $\Delta y$. 
We show the results for fixed $a=0.3$ and 
$b=0.4$, 0.6, 0.8 and 1.}
\label{f:npvy_v}
\end{figure}

We can investigate the accuracy of the binomial approximation using
the JETSET Monte Carlo (for consistency we use a fixed mass ($m=0.8$
GeV) and have no transverse momentum generation). Here we generate
2-jet ($\mbox{q}\overline{\mbox{q}}$) events and analyse the
distribution of vertices within a rapidity range, $\Delta y$. The
energy is chosen to be sufficiently large in order to avoid edge
effects from the q and $\overline{\mbox{q}}$ fragmentation
contaminating the $\Delta y$ region.  The mean, $\langle n \rangle $,
and the variance, $V$, of the resulting multiplicity distributions are
used to calculate the binomial parameters $N$ and $p$, as detailed in
appendix \ref{s:binomial}.  We will see later that binomial
distributions with these $N$ and $p$ values do indeed reproduce the
multiplicity distributions very well.  Figure (\ref{f:npvy_v}) shows
the results as a function of $\Delta y$, for various values of the
parameter $b$.  We see that for large rapidity volumes ($\Delta y
\gtrsim 5$ units) the binomial assumptions seem to work very well. 
That is to say, the observed $p$ parameter is effectively constant as
a function of $\Delta y$, whilst the parameter $N$ is linear with
$\Delta y$. This corresponds to a constant bin size $\delta
y_{\mbox{\scriptsize bin}}$. (As expected the bin size is found to be
proportional to $b m^2$.)

The behaviour of the effective JETSET $N$ and $p$ parameters at small
values of $\Delta y$ can easily be understood.  When $\Delta y$
becomes smaller than the normal bin size, $\delta y_{\mbox{\scriptsize
bin}}$, we have only one bin which is now of size $\Delta y$. In
Fig.(\ref{f:npvy_v}) all of the $N$ curves indeed tend to the limit
$N=1$. Meanwhile the observed $p$ value is the probability to find a
vertex in this single bin
\begin{equation}
p_{\mbox{\scriptsize obs}}= p \; 
\frac{\Delta y}{\delta y_{\mbox{\scriptsize bin}}}
\; \; \; \; \; (\Delta y < \delta y_{\mbox{\scriptsize bin}}) \;\;.
\end{equation}
Thus the observed $p$ becomes linear with $\Delta y$ when
$\Delta y$ is smaller than the bin size. This effect
can be seen in Fig.(\ref{f:npvy_v}).
Between the limits of large and small $\Delta y$ 
the behaviour of $N$ and $p$ are not so well determined. Here 
correlations between closely spaced vertices will play a role.

\begin{figure}[t]
  \hbox{\vbox{ \begin{center}
    \mbox{\psfig{figure=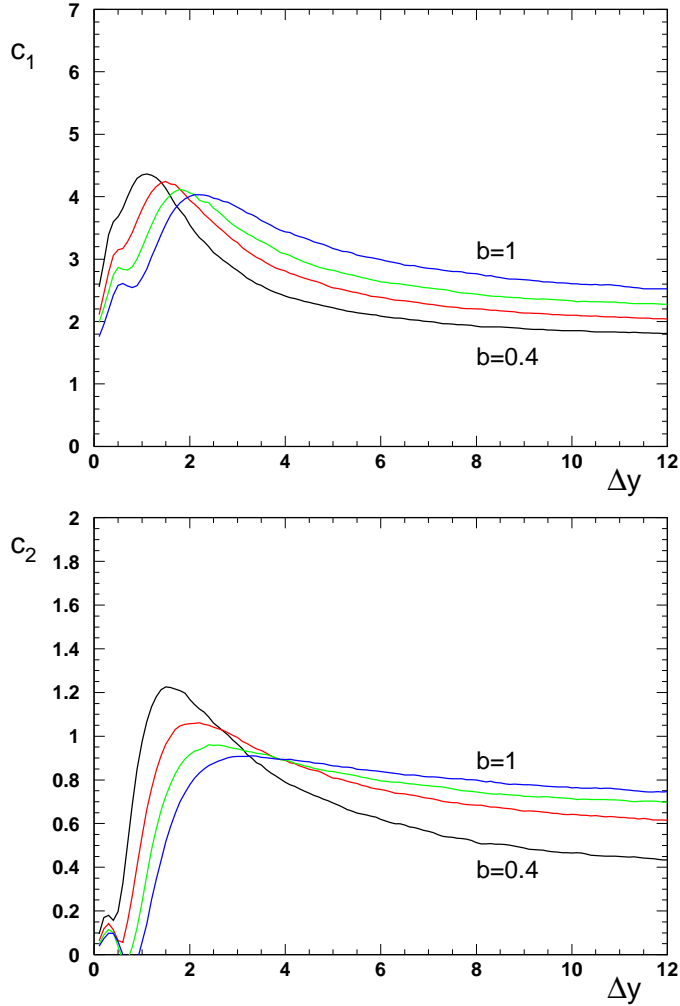,width=10cm}} 
\end{center} }}
\caption{\small The values of $c_1$ (upper plot) 
and $c_2$ (lower plot) for the vertex distributions 
produced by JETSET (for a fixed mass and 
no transverse momentum generation) 
as a function of the rapidity volume, $\Delta y$. 
We show the results for fixed $a=0.3$ and 
$b=0.4$, 0.6, 0.8 and 1.}
\label{f:c1c2}
\end{figure}

We are now in a position to 
turn our attention back to
the partition function. 
This formula should also generate a good description of the
multiplicity distribution for the vertices. We have
\begin{equation}
P_n \; = \; c_0 \;
\frac{
(c_1 \Delta y )^n
}{n!}
\exp \left ( \frac{- b m^2 \; c_2 n^2 }{\Delta y} \right )
\label{e:vgas}
\end{equation}
Here $c_0$ is a normalisation parameter and so is
determined in terms of the remaining parameters.
We can relate the parameters $c_1$ and $c_2$ to the
parameters $N$ and $p$ of the binomial distribution.
The procedure is explained in detail 
in appendix~\ref{s:lund_binomial}.
For large $\Delta y$, we obtain
\begin{eqnarray}
c_1 &=& \frac{N p}{\Delta y} \; 
\exp \left [ \frac{p}{(1-p)} \right ] \nonumber \\
c_2 &=& \frac{\Delta y}{2bm^2 N (1-p)}
\label{e:c1c2}
\end{eqnarray}
In Fig.(\ref{f:c1c2}) we show the values of
$c_1$ and $c_2$ which we obtain from our JETSET multiplicity
distributions. For large rapidity volumes, $\Delta y$, they tend to
constant values.  
We noted in section~\ref{s:lund_fwg_bo} that it is also possible to
approximate Eq.(\ref{e:vgas}) using a Gaussian distribution (with the
appropriate mean and variance). If we express the mean and variance
of Eq.(\ref{e:mult_variance}) in terms of $N$ and $p$ and solve for
$c_1$ and $c_2$, then we obtain the same expressions as
Eq.(\ref{e:c1c2}).  We note, however, that in the case of a
Gaussian distribution one has a symmetric distribution. This is not
true of either Eq.(\ref{e:vgas}) or the binomial distribution since
they both contain a term $n!$ in the denominator.

\begin{figure}[t]
  \hbox{\vbox{ \begin{center}
    \mbox{\psfig{figure=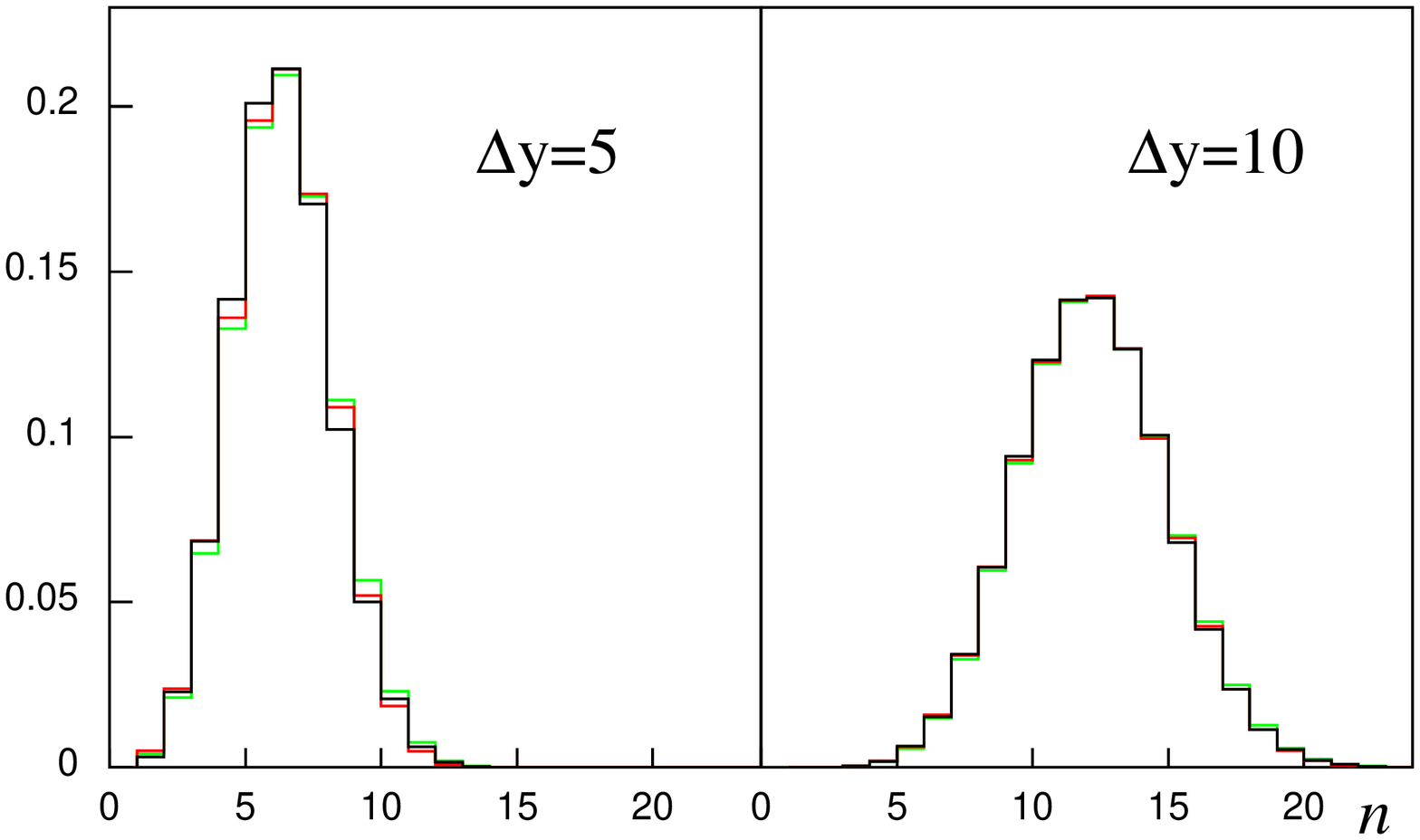,width=10cm}} 
\end{center} }}
\caption{\small The 
multiplicity distributions for the vertices
for various values of the rapidity volume $\Delta y$
as obtained from JETSET (solid curve), the binomial
distribution (medium grey curve) and Eq.(\ref{e:vgas}) (light grey curve).
}
\label{f:vertmult}
\end{figure}

Finally in Fig.(\ref{f:vertmult}) we demonstrate how well the binomial
and Eq.(\ref{e:vgas}) reproduce the observed multiplicity
distribution. We show three curves firstly the JETSET multiplicity
distribution, secondly that obtained from the binomial distribution
and finally the distribution obtained from Eq.(\ref{e:vgas}). At
$\Delta y=5$ we see very good agreement and it is difficult to
distinguish the different curves whilst at $\Delta y=10 $ all of the
curves lie on top of each other.  We thus see that the vertex
multiplicity distributions produced by JETSET do indeed agree very
well with our simple expression for the partition function, $Z_n$.

\subsection{Distribution in proper time}
\input{vert_gamma}

%% file: vert_gamma.tex
So far we have discussed the distribution of the vertices in terms of
the rapidity, $y$. If $p_T$ is neglected then the position of the
vertices is specified by one further variable $\Gamma$, which is
related to the proper time of the vertex. In this section we will
investigate how the vertices are distributed in $\Gamma$.  As we
discussed in section~\ref{s:lund}, we have for the vertices that
\begin{equation}
P(\Gamma) \propto \Gamma^a \exp(-b\Gamma)
\label{e:gamdist}
\end{equation}
which has a mean $\langle \Gamma \rangle = (1+a)/b$.
Equation~(\ref{e:gamdist}) is, however, an inclusive distribution. If
we examine vertices within a rapidity range, $\Delta y \lesssim 2$,
then we find that they are correlated. This means, for example, that
if a vertex has a large $\Gamma$ value then nearby vertices are also
likely to have large $\Gamma$ values.
\begin{figure}[t]
  \hbox{\vbox{ \begin{center}
    \mbox{\psfig{figure=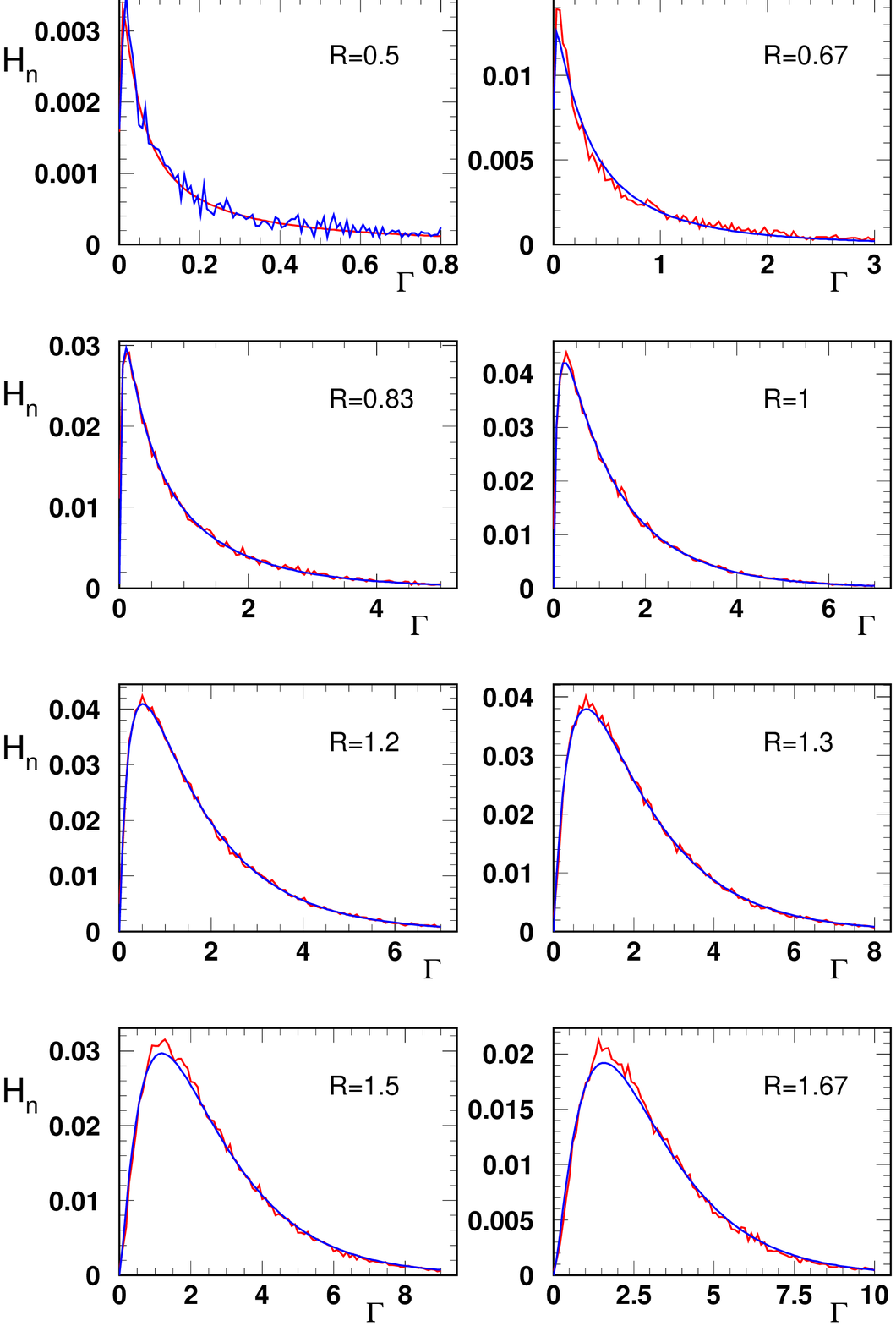,width=10cm}} 
\end{center} }}
\caption{\small 
The distributions $H_n(\Gamma,\Delta y)$ 
obtained from JETSET for the default values of the
Lund model parameters ($a=0.3$ and $b=0.58$). Also
shown are the continuous curves obtained from our fits
based on Eq.(\ref{e:h_n}). In this example
$\Delta y=6$ and $n=3 \dots 10$ . 
The corresponding $R$ values are shown on each plot. 
}
\label{f:hdist_1}
\end{figure}
\begin{figure}[t]
  \hbox{\vbox{ \begin{center}
    \mbox{\psfig{figure=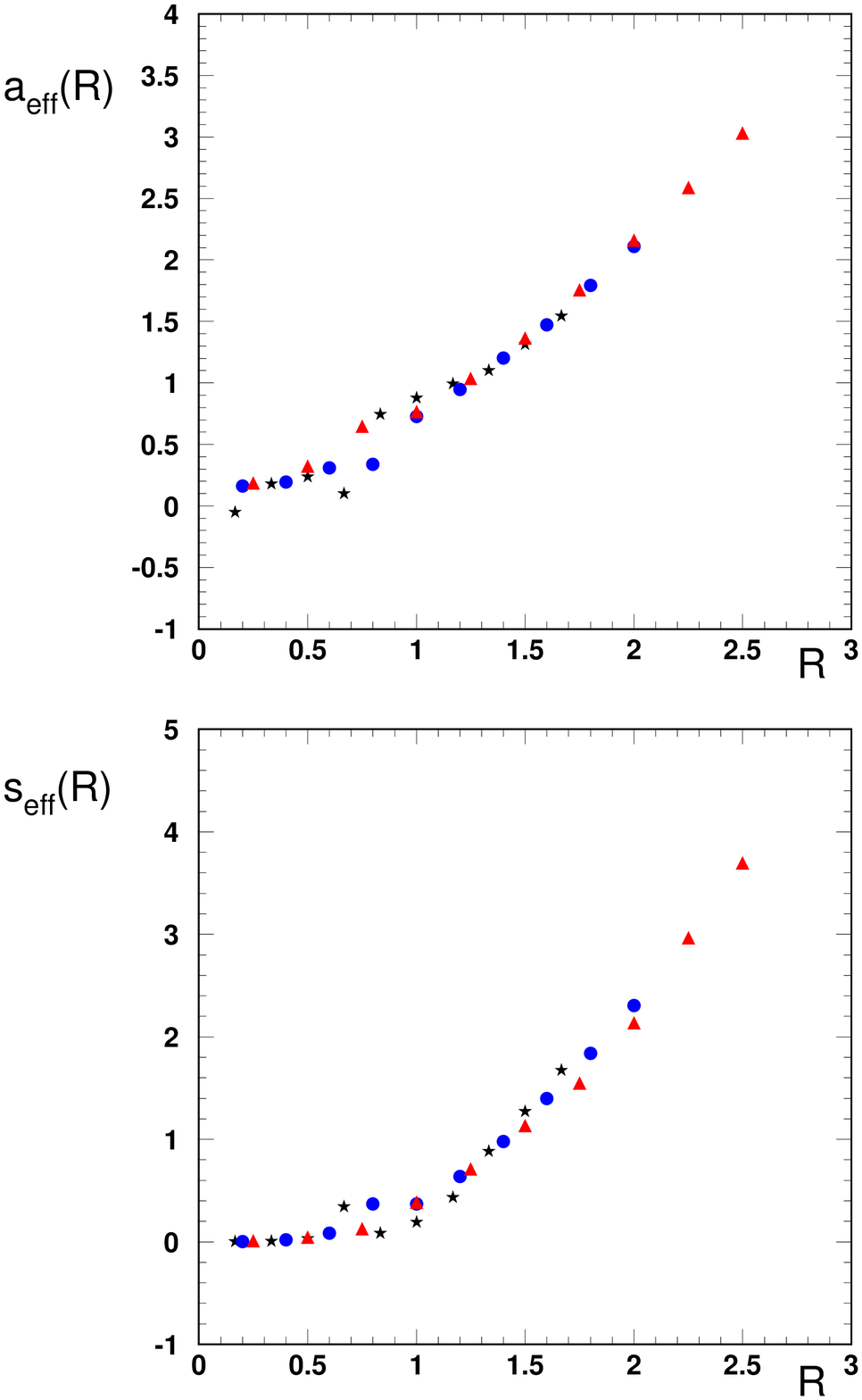,width=10cm}} 
\end{center} }}
\caption{\small 
The functions $a_{\mbox{\scriptsize eff}}(R)$ 
(upper plot) and $s_{\mbox{\scriptsize eff}}(R)$
(lower plot) versus $R$. We show the results for three different
rapidity ranges $\Delta y=4$ (triangles), $\Delta y=5$ (circles), and
$\Delta y=6$ (stars).  }
\label{f:asvr_1}
\end{figure}

We now examine how the vertices are distributed in $\Gamma$ inside a
rapidity range $\Delta y$ for various multiplicities, $n$.  Motivated
by the finite energy vertex distribution, $H(\Gamma)$, which we
considered earlier in Eq.(\ref{e:finiteH}), we parameterize the
distributions as
\begin{eqnarray}
\label{vertexsize}
H_{n}(\Gamma, \Delta y)= C \frac{\Gamma^{a_{\mbox{\scriptsize eff}}(n,\Delta y)}}
{(\Gamma + s_{\mbox{\scriptsize eff}}(n,\Delta y))^{a+1}}
\exp(-b\Gamma)
\label{e:h_n}
\end{eqnarray}
Note that taking the weighted average of $H_n$ should reproduce the
inclusive distribution of Eq.(\ref{e:gamdist}).  In
Fig.(\ref{f:hdist_1}) we show distributions in $\Gamma$ obtained from
JETSET for $\Delta y=6$.  Each plot in the figure is for a different
number of vertices, $n$, together with the corresponding fit according
to Eq.(\ref{e:h_n}).  Here the parameters $a_{\mbox{\scriptsize eff}}$
and $s_{\mbox{\scriptsize eff}}$ have been fitted for each different
$n$ value. We see that one can find values of $a_{\mbox{\scriptsize
eff}}(n,\Delta y)$ and $s_{\mbox{\scriptsize eff}}(n,\Delta y)$ for
which a very reasonable description of the $\Gamma$ distributions is
obtained. We note that the large $\Gamma$ behaviour is determined {\it
only} by the Lund parameter $b$ and not by $n$ or $\Delta y$. Thus it
is only dependent on the scale for the area law suppression. Next we
examine the dependence of both $a_{\mbox{\scriptsize eff}}$ and
$s_{\mbox{\scriptsize eff}}$ on the multiplicity and the rapidity
interval.  We have carried out fits to the $\Gamma$ distribution
obtained from JETSET for a set of values of $n$ and $\Delta y$.  We
find that both of these functions depend only on the ratio $R =
n/\Delta y$.  This can be seen clearly in Fig.(\ref{f:asvr_1}) where
we plot the results of our fits for three different values of $\Delta
y$, as a function of the density $R$.

This completes our study of the distribution of vertices produced in
the Lund model of fragmentation.  We can summarize our findings as
follows. In rapidity the vertices are approximately distributed
according to the partition function, whilst in proper time they are
distributed according to Eq.(\ref{e:h_n}). Importantly we find that
the large $\Gamma$ behaviour of the distribution in $\Gamma$ is
determined only by the area law. We also find that the functions
$a_{\mbox{\scriptsize eff}}$ and $s_{\mbox{\scriptsize eff}}$ only
depend on the density of vertices, $R$, which is itself the important
quantity in the equation of state for the gas in rapidity.

%% file: part.tex
\begin{figure}[t]
  \hbox{\vbox{ \begin{center}
    \mbox{\psfig{figure=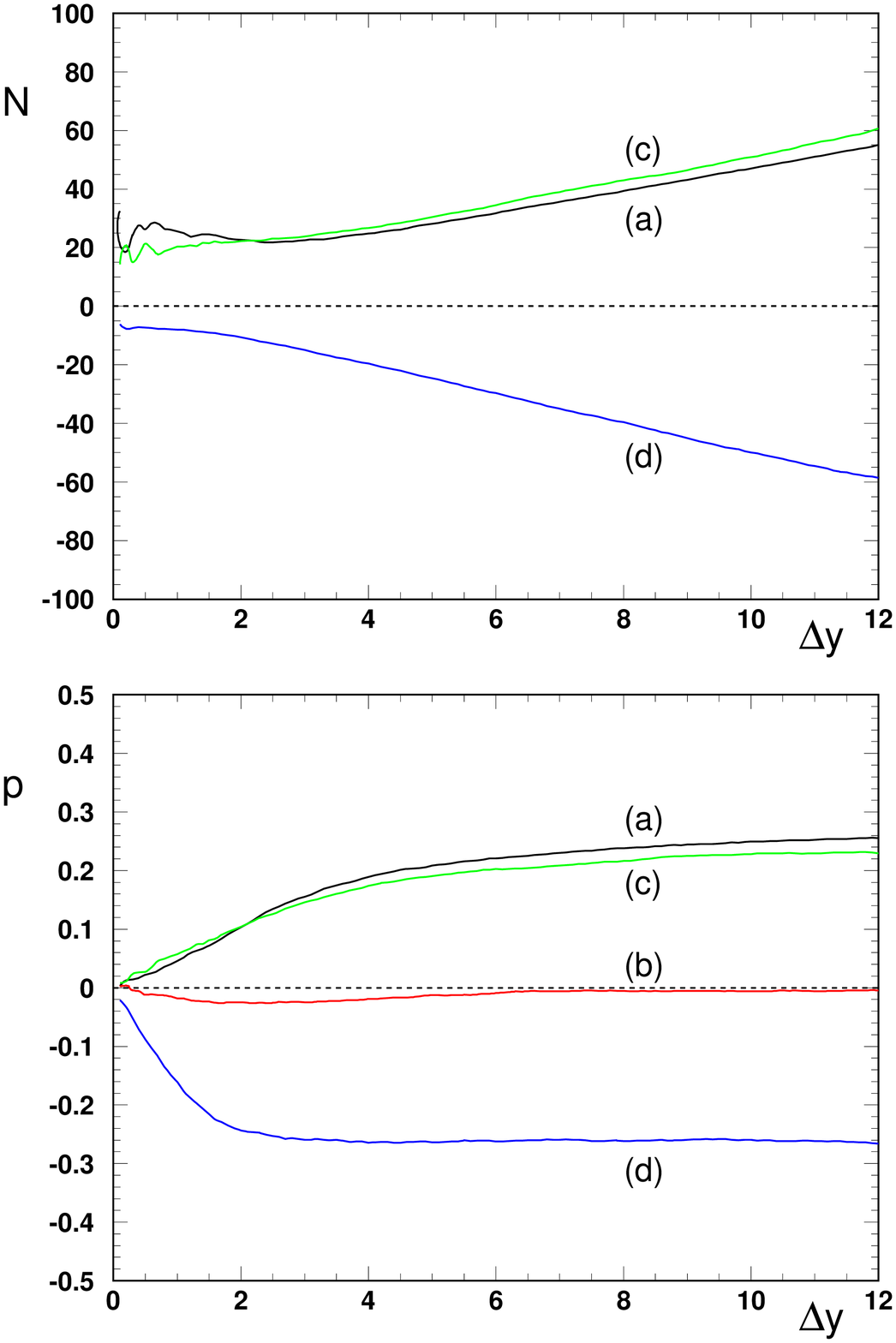,width=10cm}} 
\end{center} }}
\caption{\small 
The values of $N$ (upper plot) and $p$ (lower plot)
for the particle multiplicity distributions produced by JETSET.
The four cases shown correspond to:
(a) a single mass, no decays or $p_\perp$;
(b) complete mass spectra, but no decays or $p_\perp$; 
(c) complete mass spectra, but no decays; and
(d) complete mass spectra, charged final state particles.}
\label{f:npvy_p}
\end{figure}
For primary particles the mean multiplicity corresponds to the mean
number of vertices. The effects of going over from vertices to
particles essentially means some smearing in rapidity.  Thus the
rapidity ordering assumed for Eq.(\ref{e:Z_n}) will no longer be
true. However, for large $\Delta y$ it should still be a good
approximation.  The rapidity of a particle is distributed around the
average rapidity of the two vertices from which the particle stems
with a width of about one unit of rapidity. Therefore the particle
multiplicity distribution for a finite rapidity interval $\Delta y$
will have the same average as the vertices but a larger width. To
understand this effect we return to our simple binomial model. As
before we divide the rapidity range into $N$ equal bins with the
probability $p$ to contain a vertex.  Now we further assume that the
presence of a vertex in any bin results in a particle in one of the
two neighbouring bins with probability $q$ or in the original bin with
the remaining probability $1-2q$. In order to see how this smearing
affects the mean and the variance we compute the generating
function. The generating function for the original binomial
distribution is given by
\begin{equation}
G(z)=[1+p(z-1)]^N \;. 
\end{equation}
Some straightforward algebra then shows that 
the generating function for the above particle
distribution is given by
\begin{equation}
G(z)=[1+p(z-1)]^{(N-2)} [1+p(z-1)+p^2(z-1)^2q(1-q)]^2 \;.
\end{equation}
Thus two factors of $(1+p(z-1))$ have been modified. The mean is unchanged 
and equal to $Np$, but the variance is increased from 
$V=\langle n \rangle(1-p)$ to
\begin{equation}
V=\langle n \rangle [1-p+4pq(1-q)/N] \;.
\end{equation}
This distribution can be rather well approximated by 
another binomial distribution with
the same mean and variance. This corresponds to effective $p$ and
$N$ values
\begin{eqnarray}
p_{\mbox{\scriptsize eff}} &=& p[1-4q(1-q)/N] \nonumber \\
N_{\mbox{\scriptsize eff}} &=& \frac{N p }
{p_{\mbox{\scriptsize eff}}} \; .
\label{e:part_p_eff}
\end{eqnarray}
Thus we see how a larger spread of the particles around the vertices
(a larger $q$-value) corresponds to a larger width and a smaller
effective $p$-value. From Eq.(\ref{e:part_p_eff})
$p_{\mbox{\scriptsize eff}}$ must be larger than zero, but if we had
allowed for a spread beyond the nearest bin then negative values of
$p_{\mbox{\scriptsize eff}}$ would be possible.  This
corresponds to a negative binomial distribution. Since the bin width is
of the order of $bm_{\perp}^2$ the spread is certainly beyond
neighbouring bins in the case of pion production.

We have investigated various cases of final state production. The
multiplicity distributions can still be well approximated by binomial
distributions with constant $p$-values for large rapidity
intervals. In Fig.(\ref{f:npvy_p}) we show $N$ and $p$ as a function of
$\Delta y$ for the various cases. 

For a situation with only a single stable hadron, assumed to have the
mass $m=0.8$~GeV, and no transverse momentum generation, the result is
as expected. Comparing the multiplicity distribution with the
distribution for the vertices, we find that $p$ is decreased and $N$
is increased. The product of $N$ and $p$ is however the same for the two
distributions.

If we include the standard mixture of different hadron masses $p$ is
further reduced. We obtain in this case a distribution that is very
close to a Poissonian. Thus, as expected, the width of the
multiplicity distribution greatly increases when light pions are
produced.

Including transverse momentum generation increases $p$ to positive
values as shown in the figure. The transverse mass of the pions is
thus, in the case of the standard mixture of hadrons, not small enough
to give a negative $p$-value.

Finally, if we include the decays of unstable particles and analyse
the final charged particles then the width increases substantially and
$p$ becomes negative (corresponding to a negative binomial
distribution). Including the final uncharged particles in the analysis
results in an even more negative $p$-value.

We can summarize our findings as follows. The width of the
multiplicity distribution is very sensitive to the mass spectrum of
the produced particles. Using default JETSET the average transverse
mass is large enough to give a binomial multiplicity distribution.  In
this case the negative binomial distribution for the final state stems
from the increased width due to decays.

%% file: conclusions.tex
Inspired by the Feynman--Wilson gas analogy we have derived an
explicit form for the grand partition function of the Lund
fragmentation model. This partition function is described in terms of
the multiplicity $n$. In particular, we derive an equation of state for
the gas, corresponding to the first two terms in the virial expansion
in the particle density.

The partition function is derived in the approximation that the
particles are ordered in rapidity. This is true for the string
break-up vertices and the number of vertices corresponds to the number
of particles. Therefore, we have investigated the properties of the
partition function using the vertices. For large rapidity intervals,
we find that the average and the fluctuations of the multiplicity of
vertices are described by the partition function.

The partition function gives a multiplicity distribution which is
close to a binomial distribution. We find that the average transverse
mass of the produced particles is sufficiently large to get a
reasonable description from the approximation that the particles are
ordered in rapidity. Thus the multiplicity distribution of the
particles stemming from the string is described by an ordinary
binomial. It is the decays of the unstable particles that results in a
negative binomial distribution for the number of final charged
particles.

The distribution of the vertices for different rapidity volumes and
different multiplicities has also been investigated in terms of the
proper-time. We find that the behaviour for large proper-times is
determined only by the area-law and is independent of both the volume
and the multiplicity. For smaller proper-times the distribution is
described by a simple parametrisation. We find that the important
quantity for the parametrisation is the density of vertices in
rapidity, which in turn is described by the equation of state for the
gas.

%% file: binomial.tex
\label{s:binomial}
The binomial distribution is defined by
\begin{equation}
P(n)= 
\left ( \begin{array}{c} N \\ n \end{array} \right ) \; p^n (1-p)^{N-n} \; .
\label{e:binomial}
\end{equation}
The average, $\langle n \rangle$, and 
the variance $V= \langle n^2 \rangle -\langle n \rangle^2$ 
of this distribution are related to $N$ and $p$ via 
\begin{eqnarray}
\langle n \rangle& = & Np \nonumber \\
V &= & N p (1-p) \; .
\label{e:basicbino}
\end{eqnarray}

The binomial distributions form a family of distributions depending on
the values of $N$ and $p$. In the limit $p \rightarrow 0$ for constant
$\langle n \rangle$, the distribution becomes a Poisson
distribution. It is also possible to continue the expressions in
Eq.(\ref{e:binomial}) to negative $p$-values, which for constant
$\langle n \rangle =Np$ implies also a negative $N$. In this case the
distribution becomes a negative binomial distribution. Such a
distribution is conventionally written in the form
\begin{equation}
P_{k}(n)= 
\left ( \begin{array}{c} k+n-1 \\ k-1 \end{array} \right ) \; 
\tilde{p}^k (1-\tilde{p})^{n}
\end{equation}
where 
\begin{eqnarray}
\tilde{p} &=& \left ( \frac{1}{1-p} \right ) \; \; \; \; \;  (p<0)  \nonumber \\
k     & =& -N  \; \; \; \; \; \; \; \; ~~~~  (N<0)  \;.
\end{eqnarray}
Note that the relationships of Eq.~(\ref{e:basicbino}) for the average
and the variance remain true, even when $p$ and $N$ are both
negative. Negative $p$-values correspond to distributions which are
wider than a Poissonian.  Thus the negative binomial distributions belong
to the same larger family as the (ordinary) binomial and Poisson
distributions. Within this family the width can vary from zero to
infinity. Ordinary and negative binomials correspond to $V$ smaller
and larger than $\langle n \rangle $ respectively, with the Poisson
distribution as the limiting case in between.

%% file: lund_binomial.tex
\label{s:lund_binomial}
Our aim is to determine the $c_1$ and $c_2$ parameters
of Eq.(\ref{e:vgas}) from the $N$ and $p$ parameters of 
a binomial distribution.
We begin by using Stirlings approximation to write the
binomial distribution as 
\begin{eqnarray}
\ln(P) & =&  \frac{\ln\!N}{2}+N\ln\!N-(N-n)\ln(N-n)-n- \nonumber \\
          &  &  \frac{\ln(N-n)}{2}+N\ln(1-p)+n\ln
                \left(\frac{p}{1-p}\right) -\ln(n!)
\label{e:binomial_sterling}
\end{eqnarray}
whilst the distribution in Eq.(\ref{e:vgas}) can be written as
\begin{equation}
\ln(P_n) = \ln(c_0) - \frac{bm^2c_2n^2}{\Delta y}+n\ln(c_1\Delta y)-\ln(n!).
\label{e:gas_binomial}
\end{equation} 
We now express $n$ as $n=\langle n
\rangle + x$, where $\langle n \rangle $ is the mean.
Next we subtract Eq.(\ref{e:binomial_sterling}) from
Eq.(\ref{e:gas_binomial}) and expand around $x=0$ up to terms of order
$x^2$.  Equating the series coefficients to zero determines the
parameters $c_1$ and $c_2$ in terms of $N$ and $p$. We obtain
\begin{eqnarray}
c_1 \Delta y &=& Np \;
\exp \left [ \frac{1-2p+2Np-2Np^2}{2 N (1-p)^2} \right ] \nonumber \\
\frac{b m^2 c_2 }{\Delta y}&=& \frac{2N(1-p)-1}{4N^2(1-p)^2}
\end{eqnarray}
Which for large $\Delta y$ can be simplified to
\begin{eqnarray}
c_1 \Delta y &=& Np \; \exp \left [ \frac{p}{(1-p)} \right ] \nonumber \\
\frac{b m^2 c_2}{\Delta y} &=& \frac{1}{2 N (1-p)}
\end{eqnarray}
If we insert these expressions into  Eq.(\ref{e:vgas})
then we finally obtain
\begin{equation}
P_n \sim \frac{(Np)^n}{n!} \exp \left [ \frac{-n^2+n 2 N p}{2 N (1-p)} \right ]
\end{equation}

%% file: paper.bbl
\begin{thebibliography}{99} \vspace{-1ex}
\bibitem{r:Wilson}
K.G.~Wilson, Proc.\ Fourteenth Scottish Universities 
Summer School in Physics (1973), eds R.L.~Crawford and R.~Jennings
(Academic Press, New York, 1974). 
\bibitem{r:lund}
	\bibl{B. Andersson, G. Gustafson, G. Ingelman and T. Sj\"ostrand}
             {Phys. Rep.}{97}{1983}{31} \\
	B. Andersson, The Lund Model, ({\it Cambridge University Press}, 1998)
\bibitem{r:jetset}
	\bibl{T. Sj\"ostrand}{Comp. Phys. Comm.}{92}{1994}{74} 
\bibitem{r:bs83}
	\bibl{B. Andersson, G. Gustafson and B. S\"oderberg}{Z. Phys.}{C20}{1983}{317}
\bibitem{r:Bowler}
	\bibl{M.B. Bowler}{Z. Phys.}{C11}{1981}{169} \\
	\bibl{D.A. Morris}{Nucl. Phys}{B313}{1989}{634}
\end{thebibliography}
